\documentclass{article}
\usepackage{spconf,amsmath, graphicx,amssymb,bm, mathrsfs, array, multirow, algorithm, algorithmic, graphicx, cite, xcolor, stmaryrd, url}

\renewcommand{\bm}[1]{\ensuremath{\boldsymbol{#1}}}

\begin{document}
\title{Calibrationless MRI Reconstruction with a Plug-in Denoiser}
\name{Shen Zhao*, Lee C. Potter*, Rizwan Ahmad** \thanks{Corresponding author: Rizwan Ahmad (ahmad.46@osu.edu).}}
\address{*Department of Electrical and Computer Engineering, 
**Department of Biomedical Engineering, 
\\ 
The Ohio State University}
\maketitle

\begin{abstract}
Magnetic Resonance Imaging (MRI) is a noninvasive imaging technique that provides excellent soft-tissue contrast without using ionizing radiation. MRI's clinical application may be limited by long data acquisition time; therefore, MR image reconstruction from highly under-sampled k-space data has been an active research area. Calibrationless MRI not only enables a higher acceleration rate but also increases flexibility for sampling pattern design. To leverage non-linear machine learning priors, we pair our High-dimensional Fast Convolutional Framework (HICU) \cite{zhao2020high} with a plug-in denoiser and demonstrate its feasibility using 2D brain data.
\end{abstract}


\section{Introduction}
MRI reconstruction from highly under-sampled k-space measurements, $\mathbb{Z}$, often relies on assumptions such as sparsity of image content, smoothness of coil sensitivity map, and additive noise, etc. 
In a Bayesian framework, assumptions are expressed as priors, and the maximum a posteriori (MAP) estimation of an image, $\mathbb{X}$, and full k-space, $\mathbb{Y}$, from measurements, $\mathbb{Z}$, is  
\begin{eqnarray}
    && \max_{\mathbb{X},\mathbb{Y}} \log P (\mathbb{X},\mathbb{Y},\mathbb{Z}) \nonumber\\
    &\Longleftrightarrow& \max_{\mathbb{X},\mathbb{Y}} \log P(\mathbb{Z}|\mathbb{X},\mathbb{Y}) P(\mathbb{Y}|\mathbb{X}) P (\mathbb{X}) \nonumber\\
    &\Longleftrightarrow& \min_{\mathbb{X},\mathbb{Y}} -\log P(\mathbb{Z}|\mathbb{X},\mathbb{Y}) - \log P(\mathbb{Y}|\mathbb{X}) - \log P (\mathbb{X}) . \nonumber
\end{eqnarray}
For a regularized SENSE-based method \cite{pruessmann1999sense}, the coil sensitivity map is taken as known,  and $\mathbb{X}\mapsto \mathbb{Y}$ is a deterministic injection. The general optimization formulation of regularized SENSE is
\begin{equation}
    \min_\mathbb{X}  \mathcal{D}(\mathcal{A}(\mathbb{X}),\mathbb{Z})+\lambda \mathcal{C}(\mathbb{X})
\end{equation}
where $\mathcal{D}$ denotes a data fitting metric and $\mathcal{A}(\mathbb{X}) = \mathbb{M}\circ \mathcal{F}(\mathbb{S}\circ \mathbb{X})$ denotes composition of coil sensitivity $\mathbb{S}$, Fourier transform $\mathcal{F}$ and down-sampling $\mathbb{M}$. Here, $\circ$ denotes Hadamard product and $\lambda$ is Lagrange multiplier;  $\lambda \mathcal{C}(\mathbb{X}) + c_1$ is the negative logarithm of the image content prior, where $c_1$ is a constant. Common choices of $\mathcal{D}$ include square of Frobenius norm $\mathcal{D}(\mathcal{A}(\mathbb{X}),\mathbb{Z}) = \|\mathcal{A}(\mathbb{X}) - \mathbb{Z} \|_F^2$, which corresponds to a Gaussian noise distribution for $\mathbb{Z}|\mathcal{A}(\mathbb{X})$, or an indicator function to enforce exact data consistency
\begin{eqnarray*}
    \mathcal{D}(\mathcal{A}(\mathbb{X}),\mathbb{Z}) = \begin{cases} 0,~ \mathcal{A}(\mathbb{X}) = \mathbb{Z} \\ \infty,~\text{else.}  \end{cases}
\end{eqnarray*}
For compressed sensing methods, typically $\mathcal{C}(\mathbb{X}) = \|\mathcal{B} (\mathbb{X})\|_1$, where $\mathcal{B}$ is a sparsifying transform and $\|\cdot\|_1$ is summation of absolute values of all elements. In this case, $- \log P(\mathbb{X}) = \lambda \mathcal{C}(\mathbb{X})+c_1 = \lambda \|\mathcal{B} (\mathbb{X})\|_1 + c_1$, and the penalty term corresponds to a Laplace distribution on $\mathcal{B}(\mathbb{X})$.

Regularized GRAPPA-based methods \cite{griswold2002generalized} recover the full k-space $\mathbb{Y}$ from measurement $\mathbb{Z}$, then combine coils to map $\mathbb{Y}$ to $\mathbb{X}$. Since the coil sensitivity map is element-wise applied to each voxel through multiplication, each coil image is a modulated version of $\mathbb{X}$ and inherits a prior related to the prior on $\mathbb{X}$. For example, if $\mathbb{X}$ is $K$-sparse, then each coil image is at most $K$-sparse. The general optimization formulation of a regularized GRAPPA-based method is
\begin{equation}
    \min_{\mathbb{Y}} \mathcal{D}(\mathbb{M}\circ \mathbb{Y},\mathbb{Z})+\lambda \mathcal{L}(\mathbb{Y})
    \label{eqn: GRAPPA Like}
\end{equation}
where $\lambda$ is the Lagrange multiplier, $ \lambda \mathcal{L}(y)$ is the negative logarithm of the prior on the k-space, or equivalently on the coil images. Many physical assumptions encoded into $\mathcal{L}(\mathbb{Y})$, such as coil sensitivity smoothness or limited image support, result in a linear dependence among neighborhoods of k-space points, which in turn is equivalent to an approximate rank-deficiency of a convolution matrix $\mathcal{H}_{\mathbb{K}}(\mathbb{Y})$ built from the k-space samples, where $\mathbb{K}$ denotes the neighborhood in multi-coil k-space. Let $\bm{\sigma}(\mathcal{H}_\mathbb{K}(\mathbb{Y})) = [\sigma_1,\cdots, \sigma_n]$ denote the singular values of the convolution matrix, in non-increasing order. The squared Euclidean distance of $\mathcal{H}_{\mathbb{K}}(\mathbb{Y})$ from the closed set of matrices with rank $r$ or less is $\sum_{i > r} \sigma_i^2$, and adopting $\mathcal{L}(\mathbb{Y})=\sum_{i > r} \sigma_i^2$ \cite{haldar2014} corresponds to modeling the tail singular values as distributed according to a rectified Gaussian prior.


This work considers the problem formulation in \eqref{eqn: GRAPPA Like} with $\mathcal{D}(\cdot)$ chosen to enforce data consistency. Additionally, we assume $\mathcal{L}(\mathbb{Y}) $ of the form $ \sum_{i > r} \sigma_i^2 + \mathcal{E}(\mathbb{Y})$, where $\mathcal{E}(\mathbb{Y})$ denotes signal modeling that cannot be captured via only the rank deficiency of $\mathcal{H}_\mathbb{K}(\mathbb{Y})$. The optimization problem can be organized as: 
\begin{eqnarray}
    &&\arg \min_{\mathbb{Y}} \sum_{i > r} \sigma_i(\mathcal{H}_\mathbb{K}(\mathbb{Y}))^2 + \mathcal{E}(\mathbb{Y}) \text{ s.t. } \mathbb{M}\circ \mathbb{Y} = \mathbb{Z} \nonumber\\
    &&\Longleftrightarrow \arg \min_{\mathbb{Y}, \bm{Q}} \|\mathcal{H}_\mathbb{K}(\mathbb{Y})\bm{Q}\|_F^2 + \mathcal{E}(\mathbb{Y}) \nonumber\\
    &&\text{ s.t. } \mathbb{M}\circ \mathbb{Y} = \mathbb{Z},~\bm{Q}'\bm{Q} = I
    \label{eqn: HICU}
\end{eqnarray}
where $\mathcal{H}_\mathbb{K}(\mathbb{Y}) \in \mathbb{C}^{m \times n}$ and $\bm{Q}\in \mathbb{C}^{n \times (n-r)}$. 

The cost formulation \eqref{eqn: HICU} suggests an alternating minimization algorithm, which alternatively updates $\bm{Q}$ and $\mathbb{Y}$. For a given $\mathbb{Y}$, $\bm{Q}$ is the matrix of singular vectors corresponding to the tail singular values. For a given $\bm{Q}$, updating $\mathbb{Y}$ is a regularized least squares problem. One method for this regularized least-squares sub-problem is proximal gradient descent \cite{beck2009fast, mazumder2010spectral}, where for a function $\mathcal{F}(\mathbb{Y})$, its proximal operator is $\text{prox}_\mathcal{F}(\mathbb{Y}) = \arg \min_\mathcal{\mathbb{X}} \frac{1}{2}\|\mathbb{X}-\mathbb{Y}\|_F^2+\mathcal{F}(\mathbb{X})$. Proximal gradient descent for updating $\mathbb{Y}$ given $\bm{Q}^{(k)}$ is
\begin{eqnarray*}
\mathbb{Y}^{(k+1)} = \text{prox}_{\eta_k\mathcal{E}} \left( \mathbb{Y}^{(k)}-\eta_k\nabla_{\mathbb{Y}} \left\|\mathcal{H}_\mathbb{K}(\mathbb{Y})\bm{Q}^{(k)} \right\|_F^2\bigg |_{\mathbb{Y}=\mathbb{Y}^{(k)}} \right)
\end{eqnarray*}
where $\eta_k$ is the step size for the $k^{\text{th}}$ iteration. Moreover, the proximal operator $\text{prox}_{\eta_k\mathcal{E}}(\cdot)$ can be considered as an additive Gaussian denoising operator \cite{venkatakrishnan2013plug,ahmad2020plug}. 

In this work, we use plug-in denoisers $\mathcal{V}(\cdot)$ to replace the proximal operator, and we pair $\mathcal{V}(\cdot)$ with our High-dimensional Fast Convolutional Framework (HICU)~\cite{zhao2020high} to improve the reconstruction quality.

\section{Methods}
HICU employs the prior of linear dependency in k-space and leverages the efficiency of randomized numerical methods. The cost function in \eqref{eqn: HICU} is optimized using three numerical techniques: randomized singular value decomposition (rSVD), center-out strategy (CO), and randomized mixing of null space basis vectors inspired by Johnson Lindenstrauss lemma (JL). The CO strategy invokes shift invariance of the linear prediction property to operate on a small, low-noise, center portion of k-space at early iterations. Taken together, the numerical strategies yield both fast computation and low memory requirement. In this work, we use a denoiser $\mathcal{V}(\cdot)$ to utilize additional modeling assumptions corresponding to $\mathcal{E}(\mathbb{Y})$. Specifically, we consider two: soft thresholding in a stationary wavelet transform domain (SWT), and a trained deep neural network (DNN) denoiser. The pairing of HICU with a denoiser is given in Algorithm~\ref{alg:HICU}.
\begin{algorithm}[!ht]
\caption{HICU with a plug-in denoiser}
\label{alg:HICU}
\begin{algorithmic}[1]
    \REQUIRE 
    Zero-filled observed k-space, $\mathbb{Z}$;
    Sampling mask, $\mathbb{M}$;
    Kernel mask, $\mathbb{K}$; 
    Region of k-space, $\mathbb{R}^{(0)}$;
    Rank, $r$; 
    Compression dimension, $p$;
    Number of iterations, $I$; 
    Denoiser, $\mathcal{V}(\cdot)$;
    Initialize $\mathbb{Y}^{(0)} = \mathbb{Z}$
    \FOR{$i = 1$ \text{to} $I$}
    \STATE Compute $\bm{V}^{(i)}$, the $r$ principal right singular vectors of $\mathcal{H}_\mathbb{K}^{(i-1)} ( \mathbb{Y}^{(i-1)} )$ via rSVD \hfill{{\bf rSVD}}\\
    \STATE Compute orthonormal basis, $\bm{Q}^{(i)} \perp \bm{V}^{(i)}$ via $r$ Householder reflections
    \STATE Select region, $\mathbb{R}^{(i)}$, on which to compute valid convolution using $\mathcal{H}_\mathbb{K}^{(i)} ( \mathbb{Y}^{(i-1)} )$ \hfill{{\bf CO}}\\
    \STATE Prepare for $G_i$ descent steps, $\mathbb{W}^{(0)} = \mathbb{Y}^{(i-1)}$
    \FOR{$j = 1$ \text{to} $G_i$}
    \STATE Compress nullspace to $p$ dimensions, $\widetilde{\bm{Q}}^{(i,j)} =  \bm{Q}^{(i)} \bm{P}^{(i,j)}$, where $\bm{P}^{(i,j)}$ is i.i.d.\ normal \hfill{{\bf JL}}\\
    \STATE Calculate the gradient, $\mathbb{G}^{(j)} = \nabla_{\mathbb{W}} \left\|\mathcal{H}_\mathbb{K}^{(i)} ( \mathbb{W}^{(j-1)} )  \widetilde{\bm{Q}}^{(i,j)} \right\|_F^2,~\mathbb{G}^{(j)} = \mathbb{G}^{(j)} \circ \mathbb{M} $
    \STATE Update with step-size set by exact line search, $\mathbb{W}^{(j)} = \mathbb{W}^{(j-1)} - \eta^{(j)} \mathbb{G}^{(j)}$ \\
    \STATE Denoise $\mathbb{W}^{(j)} = \mathcal{V} (\mathbb{W}^{(j)})$ \hfill{{\bf Denosing}}\\
    \STATE Enforce data consistency $\mathbb{W}^{(j)} = \mathbb{W}^{(j)} \circ \mathbb{M}$\\
    \ENDFOR\\
    \STATE Save result of gradient steps, $\mathbb{Y}^{(i)} = \mathbb{W}^{(G_i)}$
    \ENDFOR\\
    \ENSURE Reconstructed k-space, $\mathbb{Y}^{(I)}$
\end{algorithmic}
\end{algorithm}

\section{Experiments and Results}
\begin{figure}[h]
    \centering
    \includegraphics[width = 8.5cm]{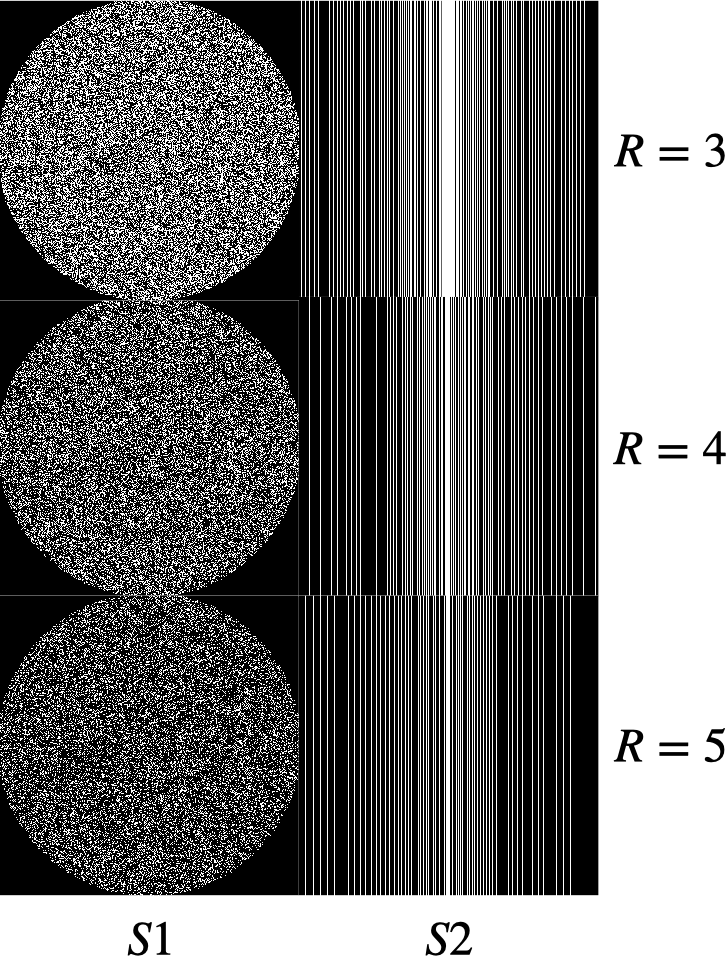}
    \caption{Sampling patterns ($S1, S2$) and acceleration rates $R=3,4,5$.}
    \label{Figure_Sampling_Pattern}
\end{figure}

We compare three algorithm choices: HICU, HICU paired with SWT (HICU+SWT), and HICU paired with DNN denoiser (HICU+DNN). A comparison of HICU to other calibrationless reconstruction methods is reported in \cite{zhao2020high}. We use $119$ T2 weighted brain datasets $D1$ to $D119$ from fastMRI~\cite{fastMRI} to benchmark the reconstruction. All datasets were fully sampled and collected in an axial orientation on 3\,T scanners using a T2-weighted turbo spin-echo sequence. Other imaging parameters included: TE 113~ms, TR 6,000---6,570~ms, TI 100~ms, field-of-view (FOV) 220~mm$ \times$ 227~mm, slice thickness 7.5~mm,  matrix size 384$\times$384, number of receive coils 16---20, and flip angle 149---180 degrees. The center slice from each dataset was used. The sampling patterns and acceleration rates are shown in Figure \ref{Figure_Sampling_Pattern}. For all three algorithm choices, rank $r = 30$, which is chosen based on one additional dataset for sampling pattern $S2, R=5$. The filter kernel size is $3 \times 3  \times N_c$, where $N_c$ is the number of coils; valid convolution is computed without the need for data padding.  CO processing is organized simply as two stages. The first uses the center $1/4 N_x \times 1/4 N_y \times N_c$ portion of k-space, $p = 8$, and iteration count  $G_i = 5$ (Step 6). The second stage uses the full k-space, $p = 32$, and iteration count $G_i = 10$. The number of outer iterations ($I$ in Step 1) dedicated to the first CO stage was tuned based on the same additional dataset to achieve fast convergence and was held constant for all six combinations of sampling patterns and acceleration rates. Computation in all cases was terminated at one minute. For computational speed, both HICU+SWT and HICU+DNN only apply the denoiser (Step 10) in the second CO stage. 

For HICU+SWT, the threshold for SWT was tuned based on the additional dataset. The threshold is proportional to the step size obtained in the exact line search (Step 9). For HICU+DNN, the DNN denoiser was trained on additional $300$ T2 weighted brain datasets with $15$\,dB additive simulated Gaussian noise. DNN network $\mathcal{N}(\cdot)$ takes a noisy image as the input and yields the estimated noise as the output. Thus, $\mathcal{V}(\cdot) = \mathcal{I}(\cdot) - \mathcal{N}(\cdot)$, where $\mathcal{I}(\cdot)$ denotes the identity map. The architecture of $\mathcal{N}$ consists of five convolution layers with ReLU function, followed by a sixth convolution layer.   The numbers of channels from the first to the sixth convolution layer are $256, 256, 128, 128, 128, 2$. Reconstruction quality is reported in terms of SNR, which is defined as $20\log_{10} (\|\hat{\mathbb{Y}}-\mathbb{Y}\|_F/\|\mathbb{Y}\|_F)$.

For all studies, processing was performed in Matlab 2020a (Mathworks, Natick, MA, US) on an Alienware Aurora Ryzen\texttrademark~desktop, equipped with an AMD Ryzen 3950X CPU and 64\,GB memory and Nvidia RTX 2080 Ti. The Matlab code for HICU is provided at \url{https://github.com/OSU-CMR/HICU}.

Figure \ref{Figure_Sampling_Pattern} shows the two sampling patterns (2D random downsampling $S1$ and 1D variable density downsampling $S2$) at three acceleration rates $R=3,4,5$.

\begin{figure}[h]
    \centering
    \includegraphics[width = 8.5cm]{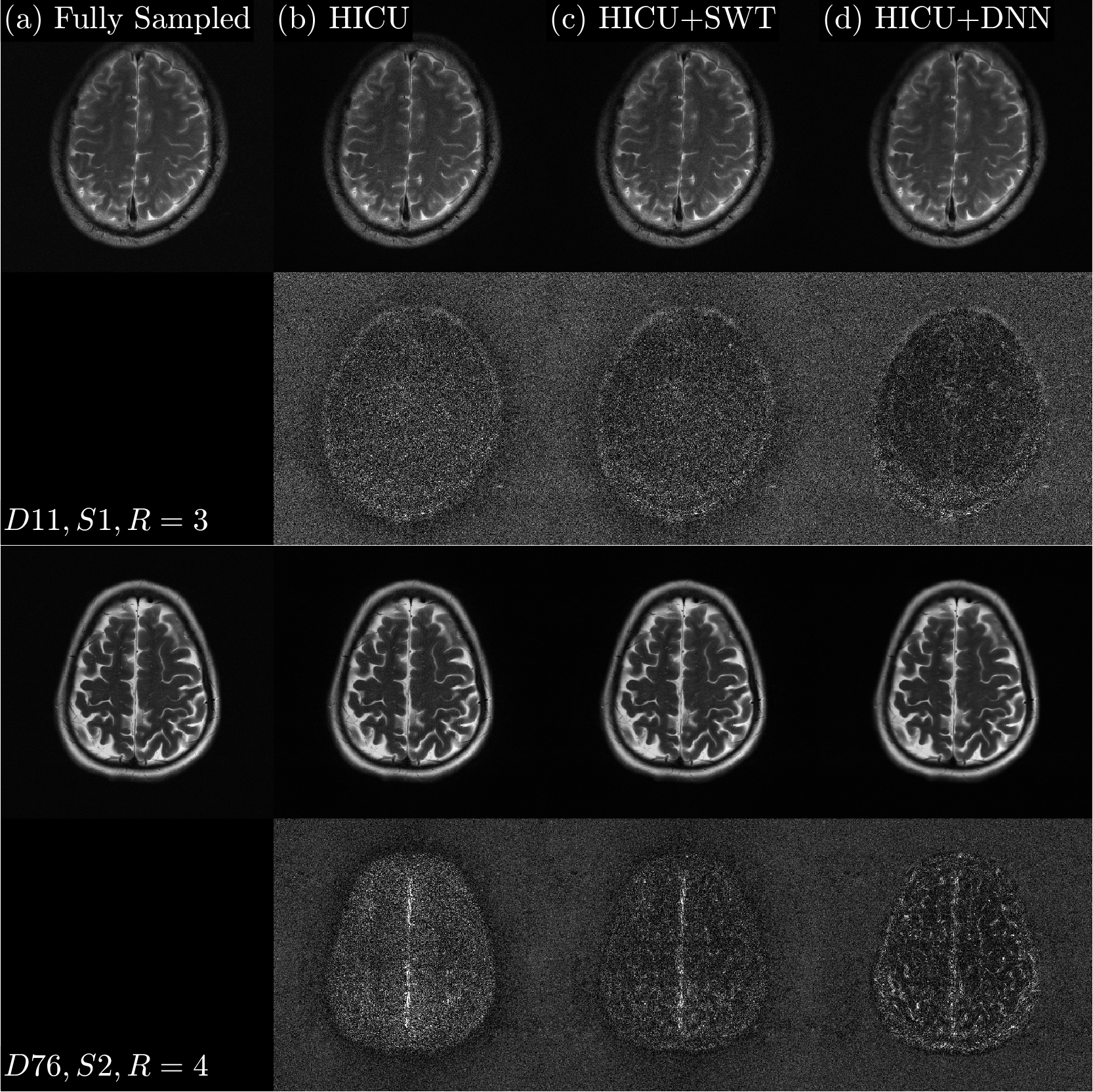}
    \caption{First and third rows: coil-combined images for each reconstruction method at one minute. Second and fourth rows: $15\,\times$\,absolute error relative to fully sampled k-space.}
    \label{Figure_2D_Images}
\end{figure}

Figure \ref{Figure_2D_Images} shows representative reconstructed images for $D11, S1, R=3$, and $D76, S2, R=4$; these examples are chosen such that the difference between SNR for HICU+DNN and HICU+SWT is closest to the average SNR difference for each sampling pattern and acceleration.

\begin{figure}[h]
    \centering
    \includegraphics[width = 8.5cm]{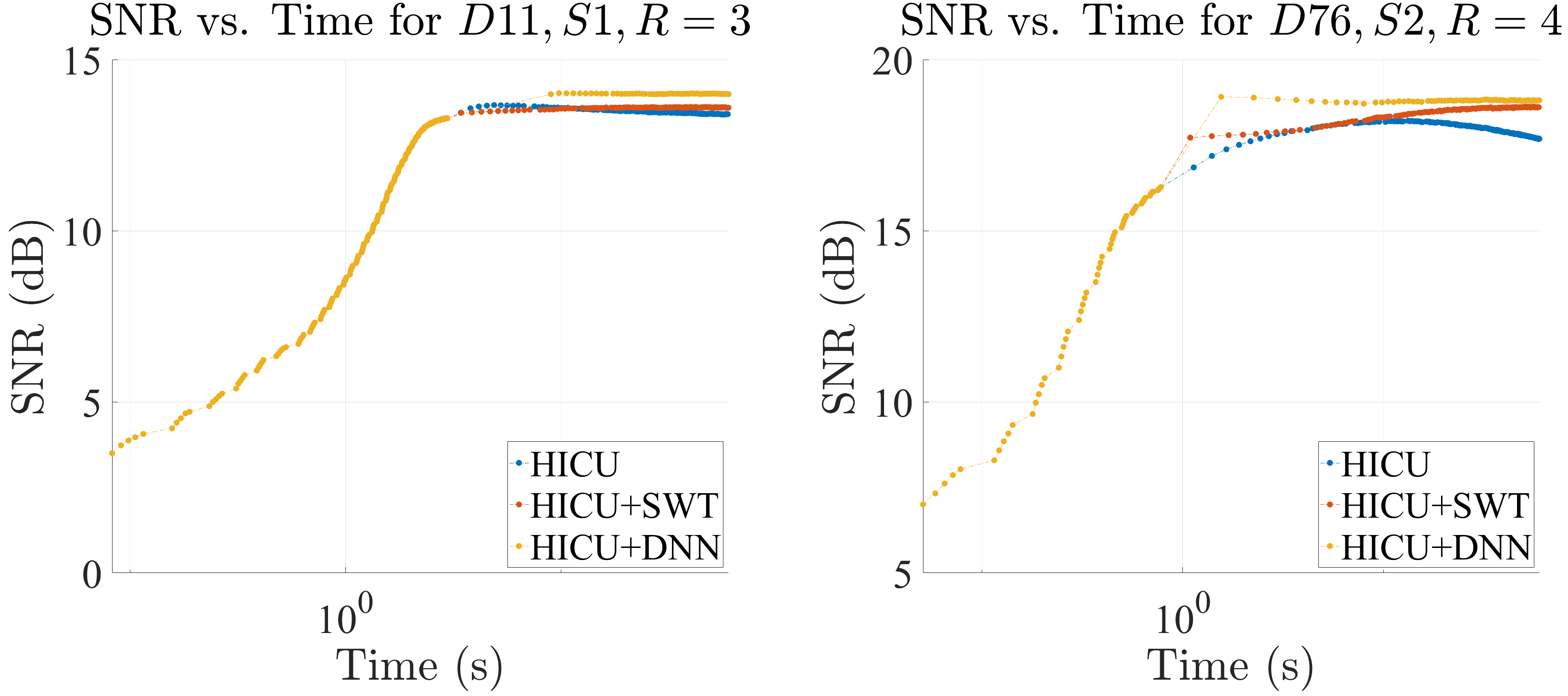}
    \caption{Reconstruction SNR versus runtime on a logarithmic scale for $D11,S1,R=3$ and $D76,S2,R=4$.}
    \label{Figure_SNR_Vs_Time}
\end{figure}

Figure \ref{Figure_SNR_Vs_Time} shows the reconstruction SNR versus log of compute time for $D11,S1,R=3$ and $D76,S2,R=4$. The use of SWT and DNN denoisers both improves peak SNR and greatly ameliorates the decay of SNR seen in HICU as iterations continue past peak SNR. 
We conjecture that reduced SNR decay for HICU+DNN, as seen in Figure~\ref{Figure_SNR_Vs_Time}, makes the performance more robust to a stopping criterion.
Although applying the plug-in denoiser requires higher floating-point operations per iteration, it provides greater SNR improvement per iteration. The net effect is that HICU+DNN achieves higher reconstruction SNR in less time than without the machine learning prior. For example, for $(S2, R=4)$, HICU needs on average $11.06$ seconds to reach peak SNR; however, HICU+DNN only needs on average $6.49$ seconds to reach the peak SNR of HICU.

\begin{figure}[h]
    \centering
    \includegraphics[width = 8.5cm]{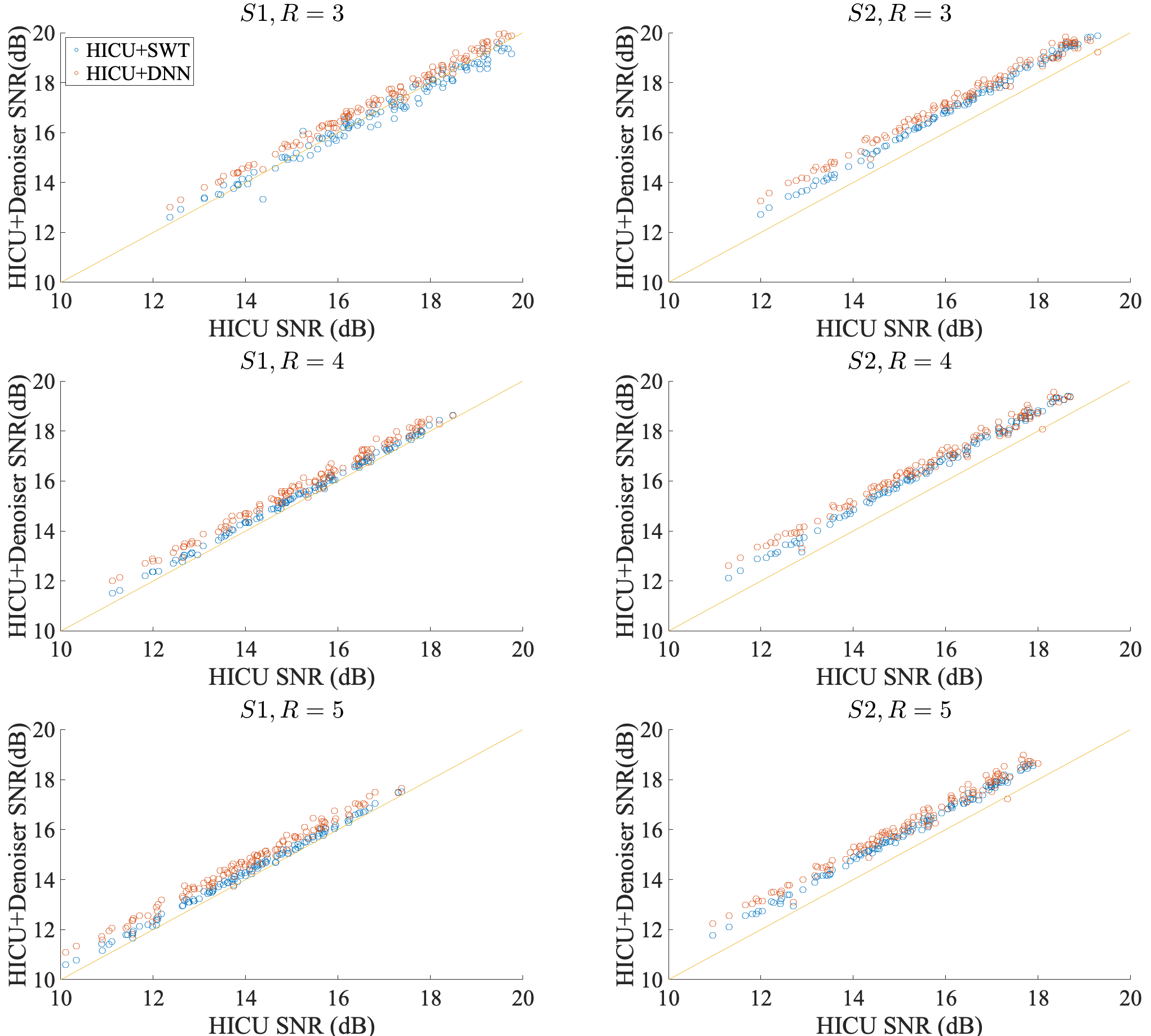}
    \caption{Comparison of reconstruction SNR (dB) between HICU and HICU-SWT as well as HICU and HICU-DNN for all sampling patterns and acceleration rates. Each dot represents a dataset.}
    \label{Figure_SNR_Methods_Data}
\end{figure}

Figure \ref{Figure_SNR_Methods_Data} shows, for each of the six combinations of sampling pattern and acceleration rate, the reconstruction SNR of HICU+SWT and HICU+DNN compared with HICU as a scatter plot of all 119 data sets. 
For all sampling patterns and acceleration rates, HICU+DNN yields consistently higher SNR than HICU without the plug-in denoiser, as seen by the cluster of red data points above the 45-degree line. The average SNR of HICU+DNN over HICU for $(S1,R=3)$, $(S1,R=4), \cdots$ to $(S2,R=6)$ are $0.39, 0.55,0.62, 1.02, 1.05, 1.00$\,dB. Moreover, DNN consistently outperforms SWT as a denoiser. For $(S1, R=3)$, HICU+SWT produces SNR similar to or even lower than HICU; this is partially due to differences between $(S1, R=3)$ and $(S2, R=5)$, which was used to tune the wavelet thresholding parameter. In addition, the DNN denoiser trained at $15$ dB can perform decently across different datasets and sampling patterns, for which the reconstruction SNR of HICU ranges from $10$\,dB to $20$\,dB.

\section{Discussion and Conclusion}
The use of a plug-in denoiser enables HICU, a fast calibrationless accelerated imaging method, to leverage an implicit machine learning prior not captured by the linear dependency image model. The denoiser is observed to improve the reconstruction quality as measured by reconstruction SNR and suppresses SNR degradation observed in HICU; additionally, despite the increase in computation per iteration, the denoiser reduces net computation time. 

\bibliographystyle{IEEEbib}
\bibliography{root.bib}

\newpage

\section{Compliance with Ethical Standards}
This research study was conducted retrospectively using human subject data made available in open access by fastMRI. Ethical approval was not required as confirmed by the license attached with the open access data.

\section{Acknowledgments}
The authors have no relevant financial or non-financial interests to disclose. This work was funded by NIH R01HL135489.
\end{document}